\documentclass[onecolumn,showpacs,floats,floatfix,superscriptaddress,aps,pra]{revtex4-1}
\usepackage{amsfonts}
\usepackage{amssymb}
\usepackage{amsmath}
\usepackage{graphicx}
\usepackage{bm}
\usepackage{color}
\usepackage{epsfig}
\usepackage{calc}
\usepackage{ifthen}
\usepackage{subfigure}
\usepackage{graphicx}
\usepackage{epstopdf}
\usepackage{epsfig}
\usepackage{braket}

\begin{document}

\title{Complete achromatic and robustness electro-optic switch between two integrated optical waveguides}

\begin{abstract}
In this paper, we present a novel design of electro-optic modulator and optical switching device, based on current integrated optics technique. The advantages of our optical switching device are broadband of input light wavelength, robustness against varying device length and operation voltages, with reference to previous design. Conforming to our results of previous paper [Huang et al, \textit{phys. lett. a}, \textbf{90}, 053837], the coupling of the waveguides has a hyperbolic-secant shape. while detuning has a sign flip at maximum coupling, we called it as with a sign flip of phase mismatch model. The a sign flip of phase mismatch model can produce complete robust population transfer. In this paper, we enhance this device to switch light intensity controllable, by tuning external electric field based on electro-optic effect.
\end{abstract}

\author{Wei Huang}
\affiliation{Engineering Product Development, Singapore University of Technology and
Design, 8 Somapah Road, 487372 Singapore}

\author{Elica Kyoseva}
\affiliation{Institute of Solid State Physics, Bulgarian Academy of Sciences, 72 Tsarigradsko Chaussee, 1784 Sofia, Bulgaria}

%\authorinfo{Further author information: (Send correspondence to Elica Kyoseva)\\Elica Kyoseva: E-mail: elkyoseva@gmail.com}

% Option to view page numbers
%\pagestyle{empty} % change to \pagestyle{plain} for page numbers   
%\setcounter{page}{301} % Set start page numbering at e.g. 301
 
\maketitle

%\keywords{integrated optics, robustness, optical switching device, electro-optic effect}

\section{Introduction}
\label{sec:intro}  % \label{} allows reference to this section

Optical switching device is the elemental device controllable by external condition, to cut-off or transmit photons, during the light transmission. Regularly, tuning of ‘switch-on’ and ‘switch-off’ is controlled by external voltages. Optical switching devices are most widely used in the optical communication \cite{Alferness1981, Farrington2011} and are also the most fundamental device for the optical computer \cite{Jahns2014, Shamir1986, Karim1992}. Furthermore, the fabrication of optical modulator can be shared the duplicated inspiration of optical switching device.

The most pristine idea of designing optical switching device is that physically and mechanically alter the light path, for example, by employing the mirrors. It is very straightforward to imagine cumbersome and slowly of this configuration. Thanks to the developments of integrated optics, nowaday, the most prevailing design of optical switching device is implemented by electro-optic effect or magneto-optic effect \cite{Shamir1986, Saleh1991}. Electro-optic (magneto-optic) optical switching device can be disciplined ‘switch-on’ and ‘switch-off’ by external voltage (magnetic field), which generates precisely controllable switching state. fast switching time and much compact integrated device. However, the fidelity of electro-optic (magneto-optic) optical switching device is awfully sensitive with the coupling length ($L_0$ in Fig.1). Due to the fabrication of integrated optics technique, it is really arduous to accurately manufacture coupling length. Therefore, fidelity of optical switching device will be drop rapidly, if there is a error in the coupling length, that makes the device is not robust against external parameters (wavelength of input light , coupling length). 

To solve the issue of robustness, a very widely used techniques to design robust optical waveguide coupler is introducing the quantum control technique. The quantum control techniques \cite{Huang17, Baum85} originally come from control quantum system from initial state to final state with fast, robustness and high fidelity. Recently, there are a large number of papers to propose quantum control techniques to various physical system, such that graphene device design \cite{Huang20171, Huang20172}, broadband half wave plate \cite{Dimova2016, Rangelov15, Huang2016} and waveguide coupler design \cite{Longhi06, Paul15, Hristova16}. In this paper, we introduce the an analytical quantum control technique, which coupling strength of the waveguides has a hyperbolic-secant shape. while detuning has a sign flip at maximum coupling, called it with a sign flip of phase mismatch model, into the optical switching device. The with a sign flip of phase mismatch model has already been illustrated in complete light intensity transfer in waveguide coupler, based on our previous paper \cite{Huang14}. In this paper, we combine the quantum control technique (phase mismatch model) and current integrated optics (electro-optic effect) to propose a novel optical switching device with robustness against varying wavelength of input light, device length and operation voltages in nanophotonic scale.

The structure of this paper is as follows. In the section II, we quickly overview of electro-optic effect and current optical switching device based on electro-optic effect. And then we introduce the phase mismatch model quantum control technique to illustrate how optical switching device of our design works in the following section. After that, we numerically investigate our optical switching robustness against varying wavelength of input light, device length and operation voltages.

\section{Electro-optic effect and current optical switching device}

In this paper, we only discuss first order electro-optic effect (well konwn as Pockels effect \cite{Saleh1991,Chmielak2011}) and high order electro-optic effect is relative excessively small, comparing with Pockels effect. Thus we just ignore the high order electro-optic effect in this configuration. Pockels effect is the material property of altering reflective index, by applying the external electric field. The general equation for the Pockels effect is given that,
\begin{equation}
n(E)= n - \dfrac{1}{2} \tau n^3 E
\end{equation} 
where $\tau$ is the electro-optic coefficient, depending on direction of the external electric field and material (in this paper, we use $LiNbO_3$ as the waveguide coupler material) and $n$ is refractive index of waveguide without external electric field.

\begin{figure}[hbtp]
\centering
\includegraphics[width=0.7\textwidth]{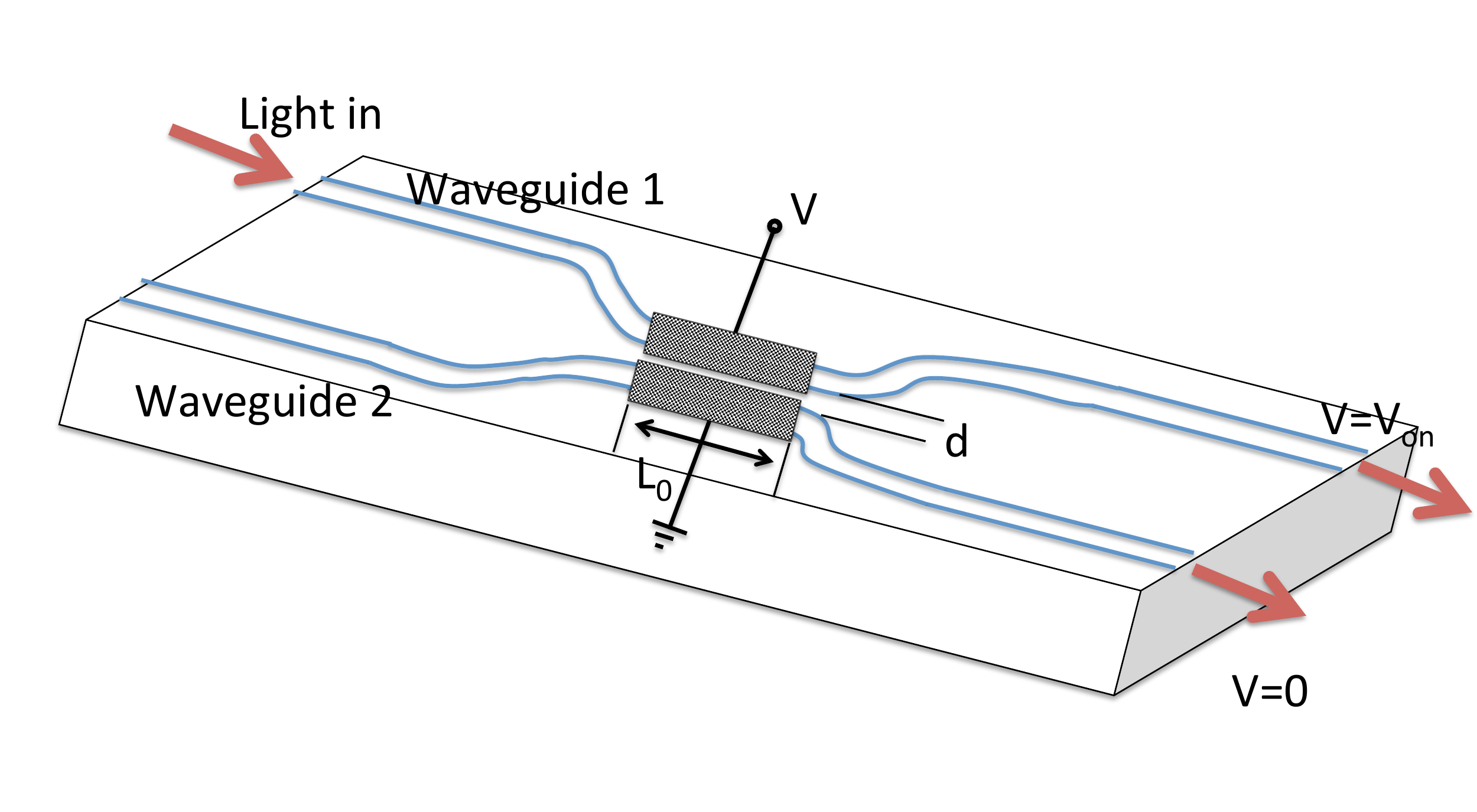}
\caption{The scheme of current integrate electro-optic optical switch. In this design, when we turn on the voltage ($V=V_{on}$), the input light is in the waveguide 1 and light out in the same waveguide. When we turn off the voltage ($V=0$), the output light is in the waveguide 2.}
\end{figure}

The scheme of previous design shows design structure in the Fig. 1 \cite{Shamir1986, Saleh1991}. Two waveguides are coupled (photons tuning from one waveguide to another waveguide) each others, due to the overlapping of the evanescent electric field, describing by coupled mode theory (CMT) \cite{Yariv1973, Haus87, Little95}. Specifically, two waveguides can be coupled, when the distance of two waveguides is closed enough. Therefore, in the Fig. 1, two waveguides only coupled at the closed areas (coupling areas) and the length of coupling areas is the coupling length $L_0$. Electrodes cover on the surface of coupling areas, by connecting to voltage $V$ and ground respectively.

For mathematical analysis, let us look inside of the coupling areas. For current design, two waveguides are parallel each others and the distance of two waveguides are the constant $d$. It is already shown that the light intensities of two parallel waveguides transfer periodically along the direction of propagation z (well known as Rabi oscillation). Assume that  coupling strength of two waveguides is $\Omega$, given by coupled mode theory \cite{Yariv1973, Haus87, Little95}, which coupling coefficient $\Omega$ exponentially depends on the distance $d$. Besides, phase matching of two waveguides is defined as $\Delta \beta = \beta_1 - \beta_2$, where $\beta_{1,2}$ are the propagation constants in waveguide 1 and 2. 

When turns off the voltage, there is no electrical field applied on the electrode of coupling area. Thus, propagation constants are the same under this circumstance and then we can get the $\Delta \beta =0$. In order to get the complete light intensity transmission from waveguide 1 to 2, the coupling length must be $L_0 = \pi / 2\Omega$, by conforming to Rabi oscillation. Therefore, the reliability of this device is much relative with   precision of the fabrication of distance $d$ and coupling length $L_0$. If these two parameters are not matching, the fidelity of ‘switch-on’ will be drop fleetly. When turns on the voltage, the external electric field implements on the electrode of coupling area, then this causes refractive index of waveguide 1 decreasing. Thus phase mismatch of detuning is no longer stay at 0, such that $\Delta \beta \neq 0$. It is already shown that \cite{Saleh1991}, the power-transfer ratio $F = P_2(L_0)/P_1(0)$, where $P_2(L_0)$ is power of waveguide 2 at $L_0$ and $P_1(0)$ is power of waveguide 1 at the beginning, given by
\begin{equation}
F=(\dfrac{\pi}{2})^2 \text{sinc}^2 {\dfrac{1}{2} [1 + (\dfrac{\Delta \beta L_0}{\pi})]^2}
\end{equation}
In terms of character of $\text{sinc}^2$ function, when $\Delta \beta L_0 = \sqrt{3} \pi$, the transfer-ratio becomes 0 and we completely block the light intensity transferring from waveguide 1 to waveguide 2, as the ‘switch-off’. The operation voltage should be, 
\begin{equation}
V_{on} = \sqrt{3}\dfrac{d}{L_0} \dfrac{\Omega \lambda_0 d}{n^3 \tau}
\end{equation} 
where $\lambda_0$ is the wavelength of input light in the air. From the Eq. (2) and (3), we can effortlessly obtain that power transfer ratio is much sensitive to coupling length $L_0$ and operation voltage $V_{on}$. Due to the fabrication process, it is very hard to control the coupling length $L_0$ very accurate. Therefore, the current optical switching device is not the robust device and troublesome to manufacture high fidelity device.

\section{The design of complete achromatic and robustness optical switch}

In this section, we propose the quantum control technique, phase mismatch model into design optical switching device. We consider two evanescent coupled waveguides based on integrate optical circuit technique. It is already shown that the evolution of the wave amplitudes of two monochromatic light beams propagating within the coupled waveguides, which the evolution can be described by a set of two coupled equations, the well known coupled mode theory \cite{Yariv1973, Haus1981},
\begin{equation}
i\dfrac{d}{dz} C(z) = H(z)C(z),
\end{equation}
where the wave amplitude vector $C(z)=[c_1(z),c_2(z)]^T$ of waveguide 1 and 2 and $I_{1,2} = |c_{1,2}(z)|^2$ are the corresponding light intensities. The operator $H(z)$ given by,
\begin{equation}
H(z)=%
\begin{bmatrix}
\beta _{1}(z) & \Omega (z) \\
\Omega (z) & \beta _{2}(z)%
\end{bmatrix}%.
\end{equation}
where $\beta_{1,2}(z)$ are propagation constants with respect to waveguide 1 and waveguide 2. The equations of $\beta_{1,2}$ can be given by $\beta_{1,2} = n_{1,2}k_0 =2 \pi n_{1,2} / \lambda_0$, where $n_{1,2}$ are the refractive indexes of waveguide 1 and waveguide 2 and $k_0$ is wave vector in the air, $\lambda_0$ is the wavelength of input light in the air. The $\Omega(z)$ is the coupling strength between waveguides. And then we define the phase mismatch of detuning as $\Delta(z) = [\beta_2(z) - \beta_1(z) ]/2$, so that we can rewrite the evolution coupled equation as,
\begin{equation}
i\dfrac{d}{dz}\left[
\begin{array}{c}
c_{1}(z) \\
c_{2}(z)%
\end{array}%
\right] =%
\begin{bmatrix}
-\Delta (z) & \Omega (z) \\
\Omega (z) & \Delta (z)%
\end{bmatrix}%
\left[
\begin{array}{c}
c_{1}(z) \\
c_{2}(z)%
\end{array}%
\right] . 
\end{equation}

Let us look at the details of the configurations of our design. In our design, two waveguides are covered by four electrodes, notation as $V_{11,12, 21, 22}$. Two electrodes, $V_{11,12}$,  cover one waveguide 1 and others two electrodes, $V_{21,22}$, cover another waveguide 2. Two electrodes $V_{11,12}$ (or $V_{21,22}$) separate half waveguide 1 (or waveguide 2) and the separation point of these two electrodes is at the maximum coupling between two waveguides (in other words, at the minimum distance between two waveguides). which the configurations of our designs see in Fig. 2. 

\begin{figure}[hbtp]
\centering
\includegraphics[width=0.6\textwidth]{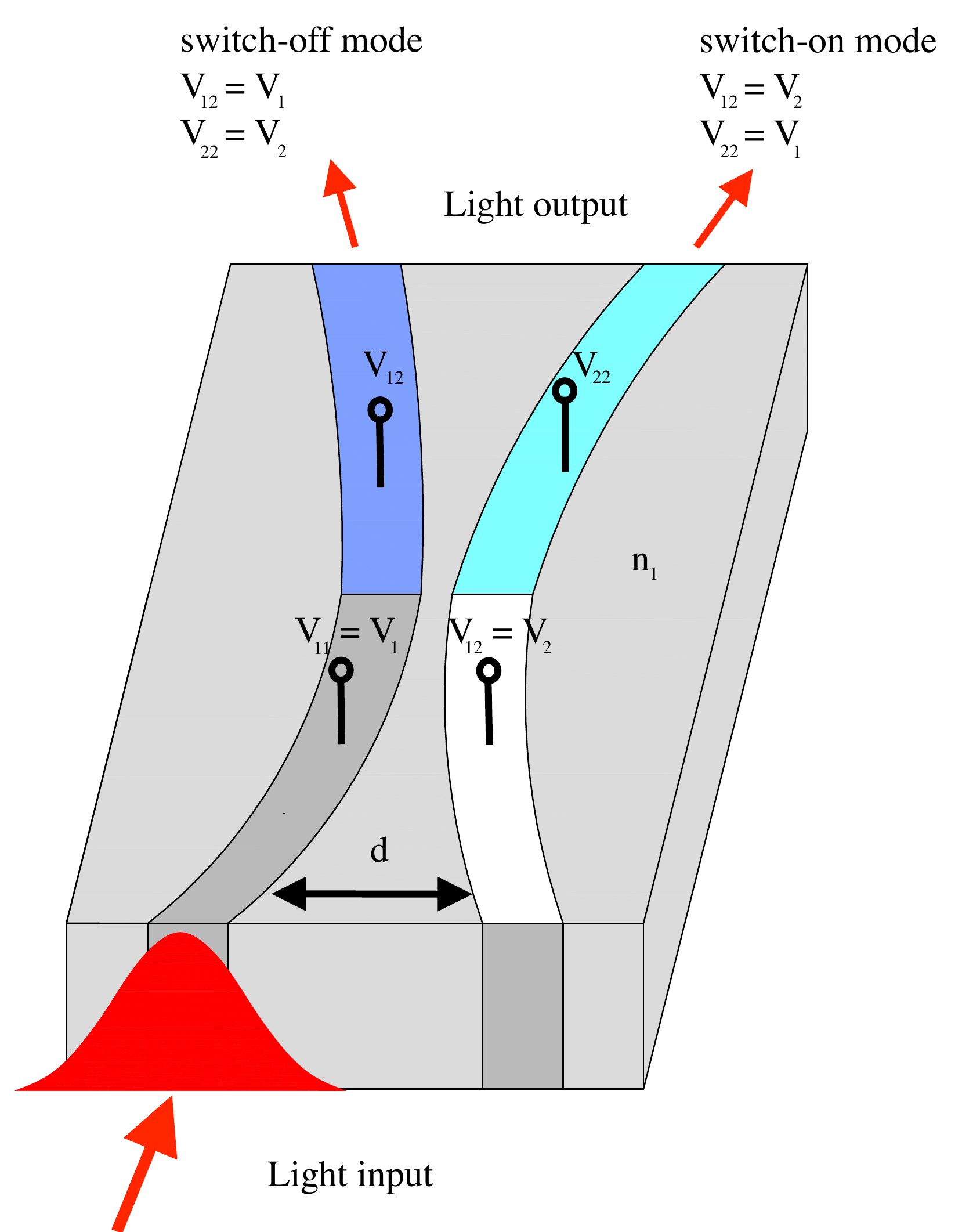}
\caption{The scheme configuration of our design, the robust electro-optic switch between two integrated optical waveguides. Our design of electo-optic switch device is controlled by employing with (or without) a sign flip of phase mismatch model to realize switch-on mode (or switch-off mode).}
\end{figure}

From the coupling mode theory, we has already obtained exponential relation between the distance of two waveguides $d$ and coupling strength $\Omega$. Based on this equation, we can engineering the coupling $\Omega(z)$, as the function of hyperbolic-secant shape, given by
\begin{equation}
\Omega(z) = \Omega_0 \text{sech}(z/L),
\end{equation}
where L is the FWHM for the coupling $\Omega(z)$, and we choose the point $z=0$ at the middle of waveguides and $\Omega_0$ is the maximum coupling coefficient.

And then according to electro-optic effect and configuration of electrodes in Fig. 2, we can get the phase mismatch of detuning $\Delta(z)$ is the function of the external electric field $E$, with
\begin{equation}
\Delta(z)=\dfrac{\beta_2(z) - \beta_1(z)}{2} = \dfrac{\pi (n_2-n_1)}{\lambda_0}= \dfrac{\pi\tau n^3( E_1 - E_2) }{2\lambda_0}.
\end{equation} 
Furthermore, we can simply take the notation as $\Delta_0 = \dfrac{\pi\tau n^3( E_1 - E_2) }{2\lambda_0}$. Now, it is very easy to find that changing the sign of phase mismatch (From $\Delta_0$ to $-\Delta_0$ or inverse) is simply by swapping the external electric fields ($E_1$ and $E_2$).  

Firstly, let us assume that we connect first half of waveguides 1 and 2 to corresponding operation voltages $V_1$ and $V_2$, with $V_{11} = V_1$ and $V_{21} = V_2$. In the ‘switch-on’ mode, the light intensity completely transfer from waveguide 1 to waveguide 2. Based on our previous research \cite{Huang14}, the with a sign flip of phase mismatch model can produce the completely transferring light intensity from one waveguide to anther waveguide. Therefore, we employ external electric fields to second half waveguide 1 and 2, by swapping voltages $V_1$ and $V_2$, comparing with default voltages setting for first half waveguide 1 and 2, such that $V_{12} = V_2$ and $V_{22} = V_1$. Thus, we can get the with a sign flip of phase mismatch model as the function of $z$, shown as

\begin{subequations}
\label{step-sech model}
\begin{eqnarray}
\Omega (z) &=&\Omega _{0}\,\text{sech}\left( z/L\right) , \\
\Delta (z) &=&\left\{
\begin{array}{c}
\Delta _{0} \\
-\Delta _{0}%
\end{array}%
\right.
\begin{array}{c}
(z<0) \\
(z>0)%
\end{array}%
.
\end{eqnarray}
\end{subequations}

The with a sign flip of phase mismatch model has already shown as completely transferring from one waveguide to another waveguide, by robustness against varying the device length and wavelength of input light. The light transfer ratio $F = P_2 (L_0) / P_1 (0)$ of with a sign flip of phase mismatch model is given by our previous study \cite{Huang14}, 
\begin{equation}
F \approx \dfrac{\Omega_0^2}{\Omega_0^2 + \Delta_0^2} [1- \dfrac{2 \Delta_0 e^{-\pi \Delta_0 L/2}}{\Omega_0} \cos(\dfrac{1}{2}\pi \Omega_0 L) + O^2]^2.
\end{equation}
From the Eq. (10), it is very easy to obtain when $\Omega_0$ is much larger than $\Delta_0$, the transfer ratio $F$ is extremely close to 1. From the coupled mode theory and Eq. (8), the maximum coupling strength $\Omega_0$ is relative with minimum distance between two waveguides and $\Delta_0$ is relative with wavelength of input light $\lambda_0$ and  electric field $E_1$ and $E_2$ (correspond to operation voltages $V_1$, $V_2$). We only make the condition satisfies $\Delta_0 \ll \Omega_0$, then complete transferring comes out. Therefore, with a sign flip of phase mismatch model completely transfers photons from one waveguide to another waveguide with robustness against varying the device length, wavelength of input light, minimum distance between two waveguides and operation voltages.

In the ‘switch-off’ mode, the photons in waveguide are completely blockaded within the waveguide 1. Based on the geometric configuration of ‘switch-on’ mode, the most simple approach is to make from with a sign flip of phase mismatch model to, we called, without a sign flip of phase mismatch model, which there is no phase mismatch appearing at the maximum coupling strength point. Establishing with the feature of the electro-optic effect, we just apply external electric fields to second half waveguide 1 and 2 as the same as the corresponding first half waveguide 1 and 2, such that $V_{12} = V_1$ and $V_{22} = V_2$. Therefore, the refraction indexes of waveguides 1 and 2 are keep the same through the device and phase mismatch of detuning $\Delta_0$ will be kept from beginning to end of device, as 

\begin{subequations}
\label{step-sech model}
\begin{eqnarray}
\Omega (z) &=&\Omega _{0}\,\text{sech}\left( z/L\right) , \\
\Delta (z) &=&\left\{
\begin{array}{c}
\Delta _{0} \\
\Delta _{0}%
\end{array}%
\right.
\begin{array}{c}
(z<0) \\
(z>0)%
\end{array}%
.
\end{eqnarray}
\end{subequations}
We can show that light completely is blockaded within waveguide 1 and there is no photon transferring from the waveguide 1 to waveguide 2 in this configuration (see Fig. 4). It is remarkable to point out the with a sign flip of phase mismatch model (in the switch-on mode) is not quantum adiabatic following, due to instantaneous changing of detuning. However, it works as almost same as the quantum adiabatic following with crossing case of detuning (complete population transfer case). If we turns our device into the switch-off mode, there is no instantaneous changing of detuning. Therefore, it is the quantum adiabatic following with no-crossing case, which produces the complete population return. Thus, we can employ the quantum adiabatic following theory to demonstrate analytical results of switch-off mode.

\section{Numerical results}

In this section, we would like to discuss and illustrate our device robustness by employing the numerically results. And then by using these results to verify how our device works with robustness in switch-on mode (see Fig. 3) and switch-off mode (see Fig. 4). After that we continue to have the discussion with coupling strength $\Omega(z)$ and detuning $\Delta(z)$ functions with Gaussian noise (see Fig. 5). 

For both the switch-on and switch-off mode, the geometric configuration and voltages setting up have already described in the section 3 (see Fig. 2). As the discussion in the section 3, let us connect first half of waveguides 1 and 2 to corresponding operation voltages $V_1$ and $V_2$, with $V_{11} = V_1$ and $V_{21} = V_2$. In the switch-on mode, second half of waveguides 1 and 2 connects the voltages with respect to $V_2$ and $V_1$, given by $V_{12} = V_2$ and $V_{22} = V_1$. Therefore, the coupling strength and detuning functions are given by equation (9). The Fig. 3 shows the results of the population transferring along with the $z$ in the switch-on mode with an example (left side of Fig.3). We install the $\Omega_0= 10$ and $\Delta_0 = 1$ ( both in the unit of $1/L$), then we assume the device length is from $z= -10$ to $z=10$ (the unit of $L$). After that we demonstrate the robustness of the $\Delta_0$ and $\Omega_0$ in the switch-on mode (right side of Fig. 3). 

\begin{figure}[hbtp]
\centering
\includegraphics[width=0.49\textwidth]{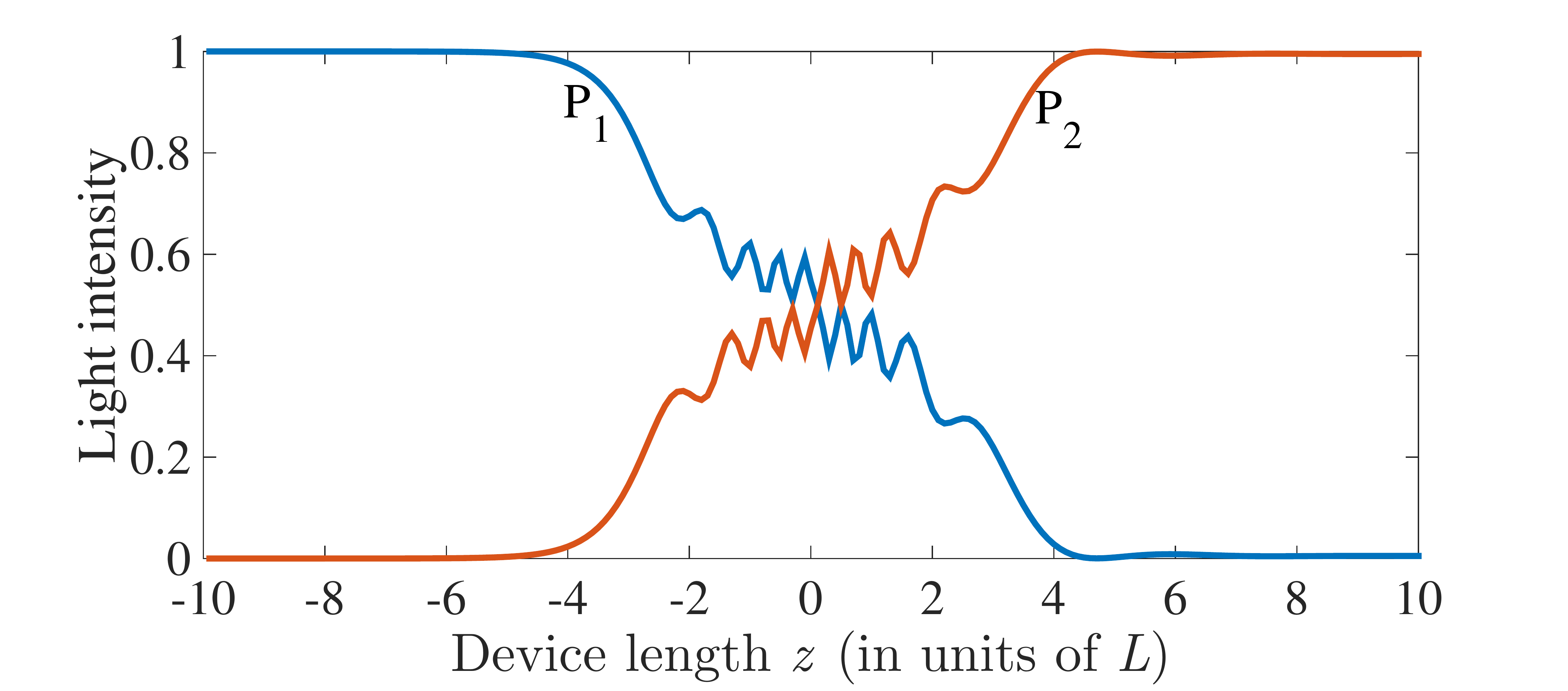}
\includegraphics[width=0.49\textwidth]{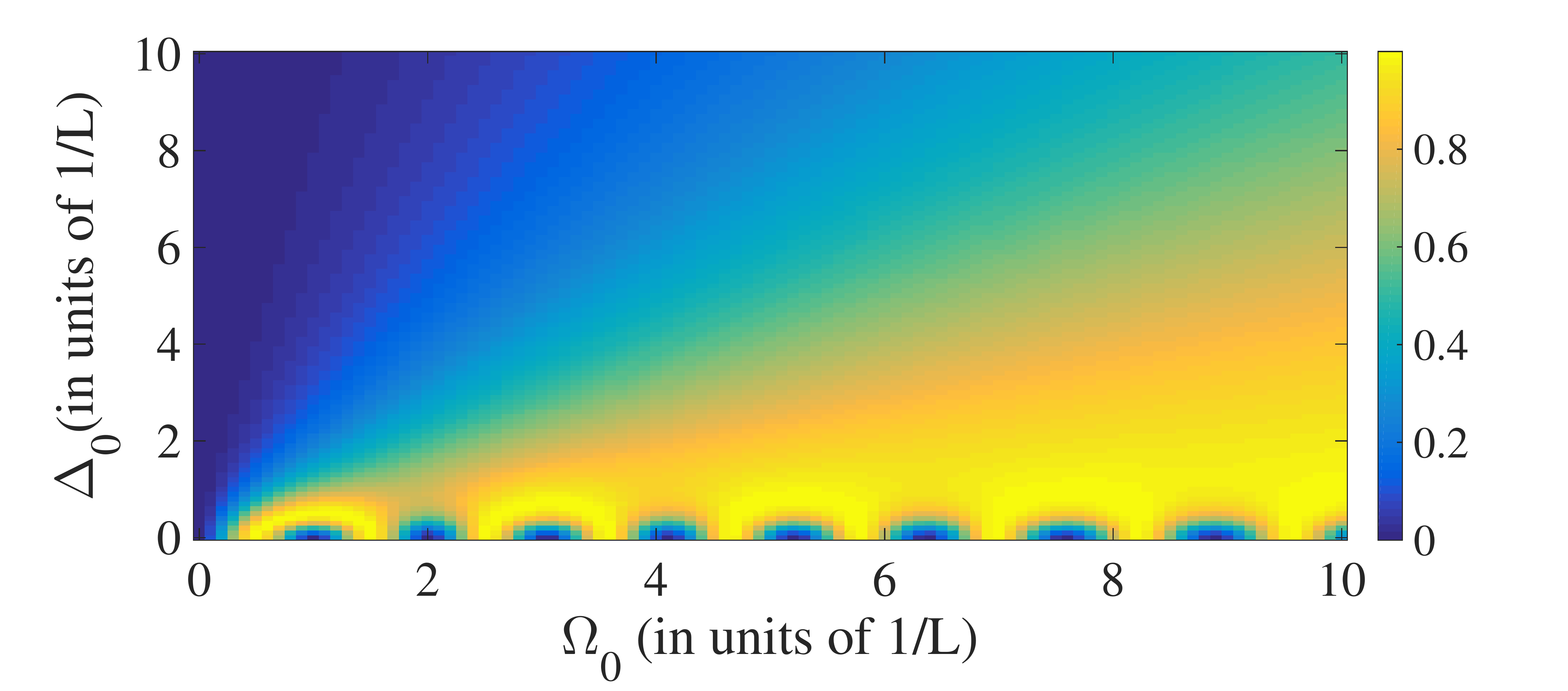}
\caption{(1) Left side: The light intensities complete transfer from waveguide 1 to waveguide 2 in the switch-on mode with an example, which is $\Omega_0= 10$ and $\Delta_0 = 1$ ( both in the unit of $1/L$). We assume the device length is from $z= -10$ to $z=10$ (the unit of $L$). (2) Right side: The robustness of coupling strength $\Omega$ and detuning $\Delta$ in the switch-on mode.  }
\end{figure}

From the results of the Fig. 3, it is very easy to notice that the configuration of switch-on delivers complete population transferring from waveguide 1 to waveguide 2 as we expected. Then For the robustness point of view, as we can obtain at right side of Fig. 3, if the condition of $\Delta_0 \ll \Omega_0$ is satisfied, there have some errors on the $\Delta_0$ and $\Omega_0$ out of our design and we still produce complete population transfer from waveguide 1 to waveguide 2. Now let us focus on switch-off mode in our design, showing the evolution of light intensities of waveguide 1 and waveguide 2 in left side of Fig. 4 and assume $\Omega_0 = 10$ and $\Delta_0 = 1$ (both in the unit of $1/L$), with device length is from $z= -10$ to $z=10$ (the unit of $L$) in the left side of Fig. 4. After that we demonstrate the robustness of coupling strength $\Omega$ and detuning $\Delta$ in the switch-off mode.

\begin{figure}[hbtp]
\centering
\includegraphics[width=0.49\textwidth]{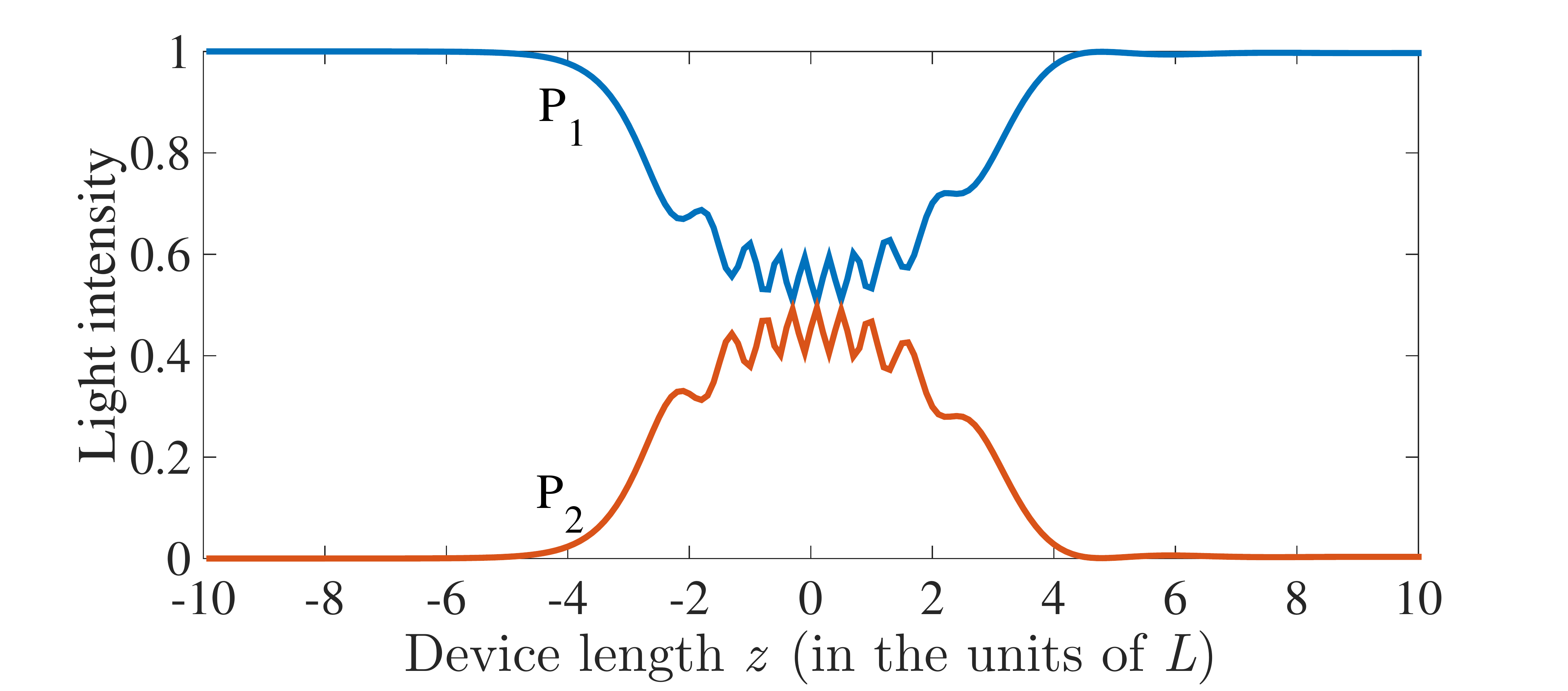}
\includegraphics[width=0.49\textwidth]{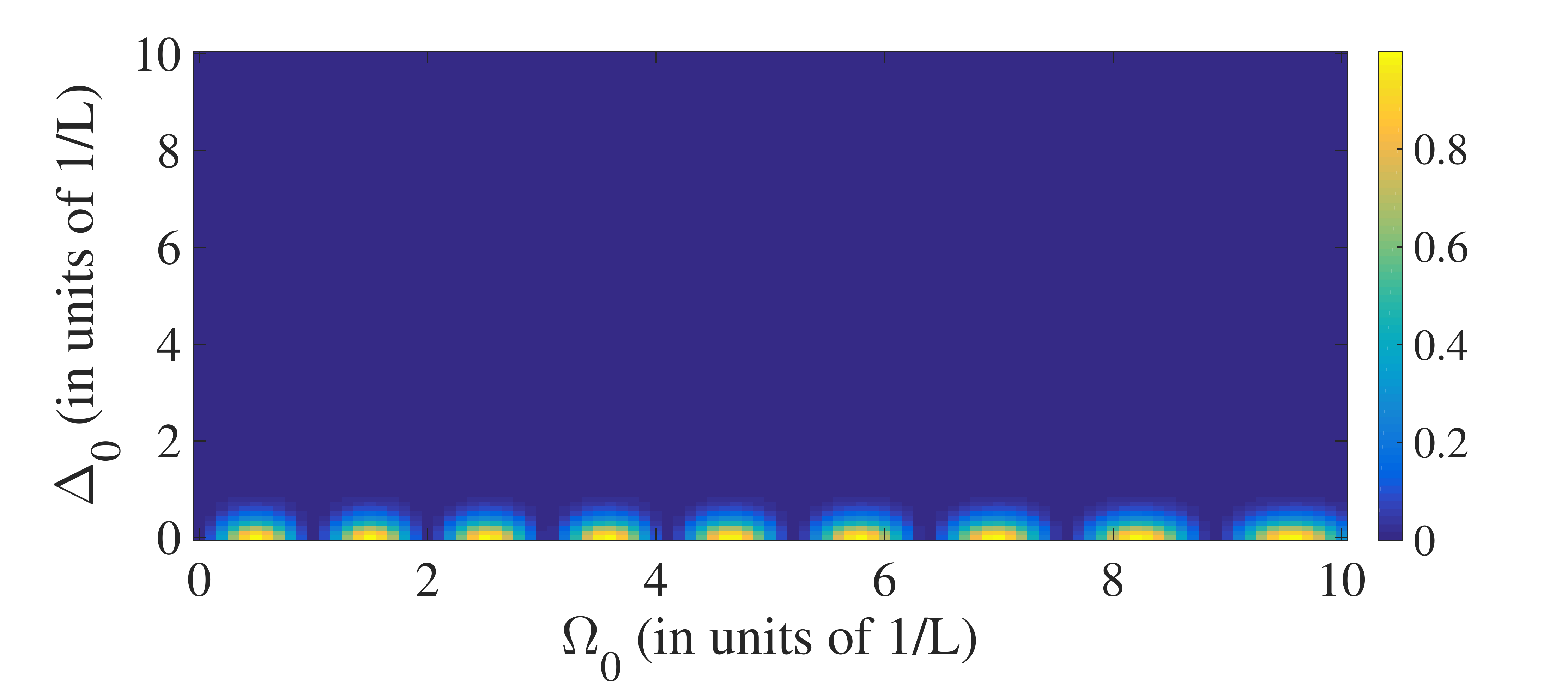}
\caption{(1) Left side: The light intensities complete return from waveguide 1 back to waveguide 1 in the switch-off mode with an example, which is $\Omega_0= 10$ and $\Delta_0 = 1$ ( both in the unit of $1/L$). We assume the device length is from $z= -10$ to $z=10$ (the unit of $L$). (2) Right side: The robustness of coupling strength $\Omega$ and detuning $\Delta$ in the switch-off mode. }
\end{figure}

As we can see the left side of Fig. 4, the light intensity of waveguide 1($P_1$) transfers from waveguide 1 to waveguide 2 and then come back to waveguide 1, so called complete population return in the switch-off mode. For the robustness of coupling strength $\Omega$ and detuning $\Delta$ in the switch-off mode in the right side of Fig. 4, it is easy to find that the operation of switch-off mode is valid at most situations, only has some small defects of robustness, when $\Delta_0$ is small (typically smaller than 0.45 $1/L$).  

At the last, we introduce the Gaussian noise into our detuning $\Delta(z)$ and coupling strength $\Omega(z)$ and then examine how this Gaussian noise influences the performance of our design. Physically, the noise of fabrication (the vibration of laser written waveguide) can be well described as Gaussian noise into the functions of detuning and coupling strength. In Fig. 5, we plot the error rate of our design, by introducing the Gaussian noise into the detuning and coupling strength functions with varying the signal noise ratio (SNR in dB). From the Fig. 5, when the Gaussian noises of detuning and coupling strength are relatively larger (SNR of Gaussian noise = 10 dB), the error rate of our design has less than 10\% (around 9\%). The error rate decreasing to less than 2\%, while the Gaussian noises are relatively low (SNR of Gaussian noise $\ge$ 17 dB). Therefore, we can conclude that our design is well robust against varying the fabrication noise. 

\begin{figure}[hbtp]
\centering
\includegraphics[width=0.7\textwidth]{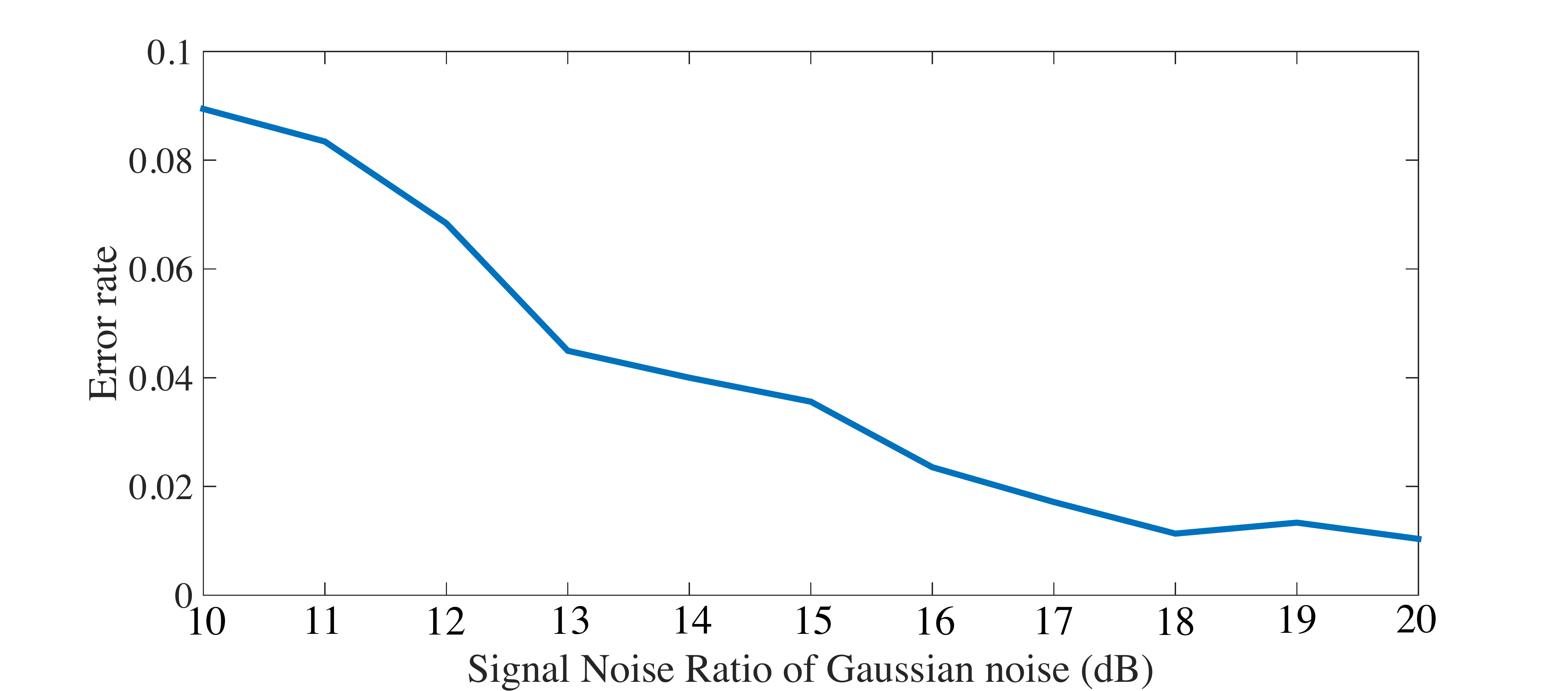}
\caption{The error rate of our design by introducing Gaussian noise to detuning $\Delta(z)$ and coupling strength $\Omega(z)$ functions, with Signal Noise Rate (SNR) from 10 dB to 20 dB.}
\end{figure}

\section{Conclusion}

In this paper, we propose a novel design for complete transfer and robust optical switching device, by employing quantum control technique (phase mismatch model) and currently integrated optics technique (electro-optic effect). Our switching operator is controlled by external voltages setting configuration, by exploiting the with (or without) a sign flip of phase mismatch model to realize switch-on mode (or switch-off mode). In this paper, we demonstrate how our design works and robustness of our optical switching device. Furthermore, we show that our optical switching device can robust against fabrication error during the fabrication processing. Therefore, we conclude that our design of optical switching device improves the robustness and performance substantially than previous design.

\acknowledgments % equivalent to \section*{ACKNOWLEDGMENTS}       
 
W. H is financially supported by Singapore University of Technology and Design (SUTD) President's Graduate Fellowship. E. K. acknowledges financial support from the European Unions Horizon 2020 research and innovation programme under the Marie Sklodowska-Curie grant agreement No 705256 - COPQE.

% References
 % bibliography data in report.bib


\begin{thebibliography}{99}
\bibitem{Alferness1981} Alferness R. C., "Guided-wave devices for optical communication." IEEE Journal of Quantum electronics, \textbf{17}(6), 946-959 (1981).

\bibitem{Farrington2011} Farrington N., et al. "Helios: a hybrid electrical/optical switch architecture for modular data centers." ACM SIGCOMM Computer Communication Review, \textbf{40}(4), 339-350 (2011).

\bibitem{Jahns2014} Jahns J., and Lee S. H., [Optical Computing Hardware: Optical Computing]. Academic press, (2014).

\bibitem{Shamir1986} Shami J., et al. "Optical computing and the Fredkin gates." Applied optics, \textbf{25}(10), 1604-1607 (1986).

\bibitem{Karim1992} Karim M. A., Awwal A. A., [Optical computing: an introduction]. John Wiley and Sons, Inc., (1992).

\bibitem{Saleh1991} Teich M. C., and Saleh, B., [Fundamentals of photonics]. Canada, Wiley Interscience, (1991).

\bibitem{Huang17} Huang W., et al. "Adiabatic following for a three-state quantum system." Optics Communications, \textbf{382}, 196-200 (2017).

\bibitem{Baum85} Baum J., Tycko R., Pines A., "Broadband and adiabatic inversion of a two-level system by phase-modulated pulses." Physical Review A, \textbf{32}(6), 3435 (1985).


\bibitem{Huang20171} Huang W., et al. "Adiabatic control of surface plasmon-polaritons in a 3-layers graphene curved configuration." Carbon, \textbf{127}, 187–192 (2017). arXiv preprint arXiv:1708.00147.

\bibitem{Huang20172} Huang W., et al. "Ultrafast electron switching device based on graphene electron waveguide coupler." arXiv preprint arXiv:1702.03748.

\bibitem{Dimova2016} Dimova E., et al. "Broadband and ultra-broadband modular half-wave plates." Optics Communications, \textbf{366}, 382-385 (2016).

\bibitem{Rangelov15} Rangelov A. A., and Kyosev E., "Broadband composite polarization rotator." Optics Communications \textbf{338}, 574-577 (2015).

\bibitem{Huang2016} Huang W., and Kyoseva E., "Application of quantum control techniques to design broadband and ultra-broadband half-wave plates." Journal of Physics: Conference Series, \textbf{752}(1), 012006 (2016).

\bibitem{Longhi06} Longhi S., "Adiabatic passage of light in coupled optical waveguides." Physical Review E, \textbf{73}(2), 026607 (2006).

\bibitem{Paul15} Paul K., and Sarma A. K., "Shortcut to adiabatic passage in a waveguide coupler with a complex-hyperbolic-secant scheme." Physical Review A, \textbf{91}(5), 053406 (2015).

\bibitem{Hristova16} Hristova S., et al. "Adiabatic three-waveguide coupler." Physical Review A, \textbf{93}(3), 033802 (2016).

\bibitem{Huang14} Huang W., Rangelov A. A., and Kyoseva E., "Complete achromatic optical switching between two waveguides with a sign flip of the phase mismatch." Physical Review A, \textbf{90}(5), 053837 (2014).



\bibitem{Chmielak2011} Chmielak B., et al. "Pockels effect based fully integrated, strained silicon electro-optic modulator." Optics express, \textbf{19}(18), 17212-17219 (2011).

\bibitem{Yariv1973} Yariv A., "Coupled-mode theory for guided-wave optics." IEEE Journal of Quantum Electronics, \textbf{9}(9), 919-933 (1973).
 
\bibitem{Haus87} Haus H., et al. "Coupled-mode theory of optical waveguides." Journal of Lightwave Technology, \textbf{5}(1), 16-23 (1987).

\bibitem{Little95} Little B. E., and Huang W. P., "Coupled-mode theory for optical waveguides." Progress In Electromagnetics Research, \textbf{10}, 217-270 (1995).
 
\bibitem{Haus1981} Haus H., and Fonstad C., "Three-waveguide couplers for improved sampling and filtering." IEEE Journal of Quantum Electronics, \textbf{17}(12), 2321-2325 (1981).


\end{thebibliography}
\end{document}